\documentclass[aps,prl,twocolumn,showpacs,superscriptaddress,groupedaddress]{revtex4}  
\usepackage{graphicx}  
\usepackage{dcolumn}   
\usepackage{bm}        
\usepackage{amssymb}   
\usepackage{amsmath} 

\def\ket#1{ | #1 \rangle}

\def\pd2v#1#2#3{\frac{\partial^2 #1}{\partial #2 \partial #3}}

\def \2x2mat#1#2#3#4{
\left( \begin{array}{cc}
#1 &  #2 \\  #3 &  #4
\end{array} \right)
}

\begin{document}

\widetext


\title{Single-photon entanglement generation by wavefront shaping in a multiple-scattering medium}

\author{Hugo Defienne }
\affiliation{ Institut Langevin, UMR787 CNRS et ESPCI ParisTech, 1 rue Jussieu, 75005 Paris, France}
\author{ Marco Barbieri }
\affiliation{Clarendon Laboratory, Department of Physics, University of Oxford, OX1 3PU, United Kingdom}
\author{ Benoit Chalopin }
\author{ Beatrice Chatel }
\affiliation{ Laboratoire Collisions, Agr\'egats, R\'eactivit\'e, (CNRS UMR 5589), IRSAMC, Universit\'e Paul Sabatier, 31062, Toulouse, France}
\author{ Ian A. Walmsley }
\author{ Brian J. Smith}
\affiliation{Clarendon Laboratory, Department of Physics, University of Oxford, OX1 3PU, United Kingdom}
\author{ Sylvain Gigan}
\affiliation{ Institut Langevin, UMR787 CNRS et ESPCI ParisTech, 1 rue Jussieu, 75005 Paris, France}

\date{\today}

\begin{abstract}
We demonstrate the control of entanglement of a single photon between several spatial modes propagating through a strongly scattering medium. Measurement of the scattering matrix allows the wavefront of the photon to be shaped to compensate the distortions induced by multiple scattering events. The photon can thus be directed coherently to a single or multi-mode output. Using this approach we show how entanglement across different modes can be manipulated despite the enormous wavefront disturbance caused by the scattering medium.  
\end{abstract}

\maketitle

Random walks describe the evolution of a system on a discrete graph where the direction of each new step is decided by tossing a coin. This simple idea describes several important physical phenomena, from localization of electrons in solids to reptation of proteins in liquids. Further, it  has diverse applications in computer science and network analysis. Quantum random walks (QRW) refer to  coherent superpositions of different trajectories of a quantum system. QRW have deep implications for quantum computing \cite{Ambainis07, Farhi08, Potocek09}, including the possibility of universal computing \cite{Childs09, Childs13}.  

Photonics has proven a versatile architecture for the implementation of quantum walks. Further, the wave-like character is easily accessible in ambient conditions, and the high degree of coherence and lack of interaction between photons in a linear optical system enable complex graphs to be implemented in the laboratory~\cite{Perets08, Schreiber10, Broome10}. Demonstrations range from multiparticle~\cite{Sansoni12}, and multidimensional walks~\cite{Schreiber12}, to analogues of Anderson localisation in condensed matter systems~\cite{Crespi13, DiGiuseppe13}, to the implementation of the boson sampling problem \cite{Broome13, Spring13, Tillmann13, Crespi13a}. Waveguide structures such as those employed in the aforementioned implementations have accessed up to 20 modes~\cite{Peruzzo10}.

Disordered multiply scattering media, such as a thick layer of densily packed dielectric particles, that scatters light in a complex manner, are highly multimode converters for light. They have been investigated in the context of quantum optics and quantum information, both theoretically ~\cite{Lohdahl05, Cande13, Cherroret11} and experimentally ~\cite{Peeters10, Beenakker09, Smolka09}. Thanks to the linearity of the elastic scattering processes, coherence is maintained during multiple scattering but the modes become highly complex, creating a so-called speckle pattern.  Recently, wavefront shaping has appeared as a powerful technique to control light propagation through such media~\cite{Vellekoop07, Mosk12}, including the ability to guide light, focus and image. Through direct characterization of the scattering matrix (SM) of a material, one can link input to output modes of the scattering medium~\cite{Popoff10}.

In this work, we explore the possibility for using such multiple-scattering media for quantum information processing, allowing access to a highly multimode system, well beyond what is possible using waveguides. A crucial point is to demonstrate the control of the quantum light travelling through and out of the medium. Recently, an iterative optimization algorithm was used to control the wavefront of a single photon and deliver it efficiently to a given position after propagation in a scattering medium~\cite{Huisman13}. In this letter, we investigate the propagation of a single photon in a scattering medium with a fully characterized SM. We show that, similarly~\cite{Huisman13}, we can control the incident wavefront of the single photon and deliver it to a given position, but do not require any iterative feedback to do so. Furthermore, we are able to experimentally address a second critical aspect of quantum propagation, namely that the coherence between different paths of the photon through the medium is preserved. This allows us to demonstrate the controlled manipulation of an entangled state between two output modes of a single photon after propagation in multiple-scattering medium.

\begin{figure}
\includegraphics[scale=1]{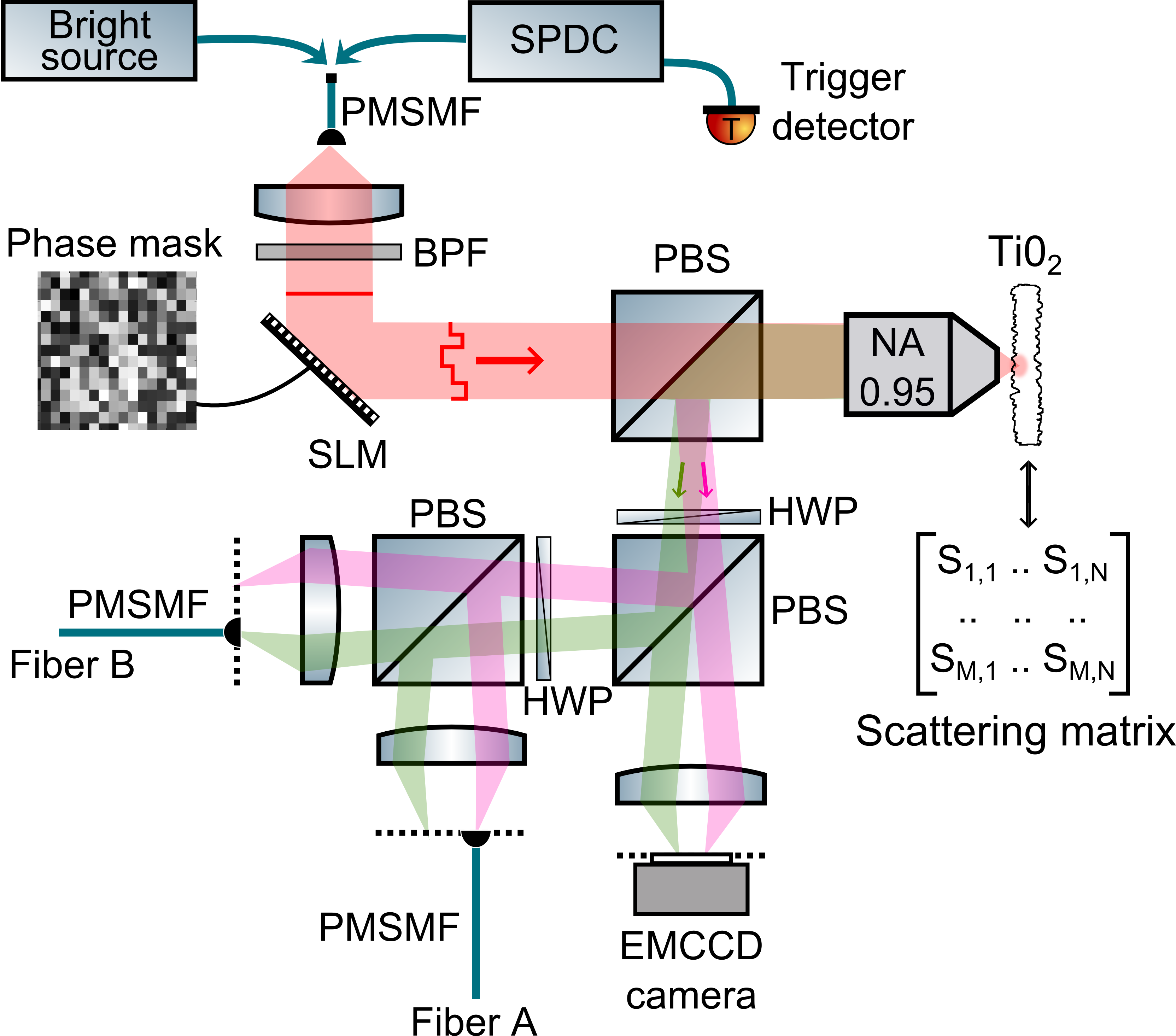}
\caption{\label{fig:setup} Apparatus for control of the propagation of non-classical light through a scattering medium. The setup consists of two light sources and an optical wavefront shaping setup. The bright classical source is used to characterize the scattering matrix (SM) of the medium. After this step, we are able control the propagation through the scattering medium of heralded single-photons generated by a non-classical source (SPDC). Single-photon focusing and single-photon entangled state generation is achieved using a Spatial Light Modulator (SLM) displaying the appropriate phase pattern determined using the SM. The scheme drawn here represents a particular case where the phase pattern displayed on the SLM focus the scattered light in two output modes (Purple and green rays). This type of SLM phase pattern is used for single-photon entangled state generation.\\
BPF: BandPass Filter ; PBS: Polarisation Beam Splitter ; PMSMF: Polarisation Maintaining Single Mode Fiber ; HWP: Half Wave Plate ; SLM: phase only Spatial Light Modulator ; EMCCD: Electron Multiplied Charged Coupled Device camera. }
\end{figure}

The experimental setup is shown in Fig.~\ref{fig:setup}. Light from a polarisation maintaining single mode fiber (PMSMF) is collimated and reflected by a phase-only Spatial Light Modulator (SLM). The shaped wavefront is then focused into the scattering medium using a microscope objective (MO) with a numerical aperture of 0.95. The scattering medium is an opaque layer of TiO$_2$ with a mean free path $l^* \thickapprox 1\mu m$ to $2\mu m$. The same MO is used to collect backscattered light from the medium. This configuration gives a higher collection efficiency of the scattered light and allows a broader effective bandwidth of the medium than is possible in transmission.  Orthogonal polarisation of the input and output fields are selected using a polarizing beam splitter (PBS) to filter out single scattering events and thus collect only light coming from multiple scattering events. The light is then collected either with an (a) electron multiplied charged coupled device camera (EMCCD) or with (b) two PMSMFs. The amplitude ratio between each paths is controlled combining a half wave plate (HWP) with a PBS. In the first path the output plane of the medium is imaged on the EMCCD with a lens of  focal length $f=150$mm. In the other path the output field is imaged onto two conjugated planes using two lenses of $f=30$mm, each containing one of the two output PMSMFs mounted on translation stages. Both fibers and the EMCCD camera are positioned in optical planes conjugate to the surface of the scatterer. The imaging system is optimized to match the speckle grain diameter to the mode field diameter of the fibers ($5.3\mu m$). 

The experiment is realized in two steps. First, the SM of the linear system is measured using a classical source, a continuous wavelength infrared diode laser with a central wavelength of 814 nm. This laser works at very low intensity, below threshold, so the light output has a broad spectrum (FWHM $\simeq 13$ nm). A 3nm band around 810nm is selected using a narrowband filter. Since the dwell time in the medium is short for reflection, the speckle contrast of the reflected light is close to unity~\cite{Curry11}, thus ensuring that light propagation can be considered as quasi-monochromatic. This is in contrast with a transmission experiment where narrower medium bandwidth and pulse broadening can be observed~\cite{Mccabe11}.  The SM ($S$) of an optical system is a $M \times N $ matrix that connects $N$ modes of an input field to $M$ modes of an output field

\begin{equation}
E^{out}=S \times E^{in}
\label{matrix} 
\end{equation}

where $ \times $ is a matrix multiplication. A method to measure the SM is described in~\cite{Popoff10}. The projection $E_m^{out}$ of the output field on the \textit{mth} output mode is given by $E_m^{out} = \sum_n S_ {mn} E_n^{in} $. Here, the SM has been determined using an input field decomposed into $N=1024$ input modes, corresponding to $32 \times 32$ macro-pixels of the SLM. At the output we measure simultaneously the SM relative to the $512 \times 512$ EMCCD pixels and the two output PMSMFs ($M=512 \times 512+2$) . By inverting equation~(\ref{matrix}), we can determine the input field corresponding to a given output field.  The input field is then shaped correctly by programming the SLM accordingly to generate the desired output field. In our experiment the hermitian conjugate $S^\dag$ of the matrix, corresponding to phase conjugation~\cite{Popoff10}, is used rather than the inverse $S^{-1}$. Moreover, the SLM used here shapes only the phase of the input field and not its amplitude. Despite these two experimental details, output light can be efficiently controlled and so focused in one or more chosen output modes.

Once the SM is measured, the classical source is replaced by a non-classical heralded single-photon source. By ensuring that the non-classical light is in the same optical mode (spectral, spatial and polarization) as the classical light for which the SM has been measured, the SM can be used to control the single photon propagation.  A periodically poled potassium titanyl phosphate (PPKTP) crystal pumped with a 50 mW continuous wavelength laser diode at 405 nm produces pairs of photons by a  type-II Spontaneous Parametric Down Conversion (SPDC) process. Both photons are filtered using the same narrowband filter centered at 810 nm (FWHM $\simeq 3$ nm) as the classical source, yielding a coincidence rate of up to $10^5 s^{-1}$. One photon of the pair is used as a trigger photon measured with a detector T heralding its twin which is injected in a PMSMF. 

\begin{figure}
\includegraphics[scale=0.89]{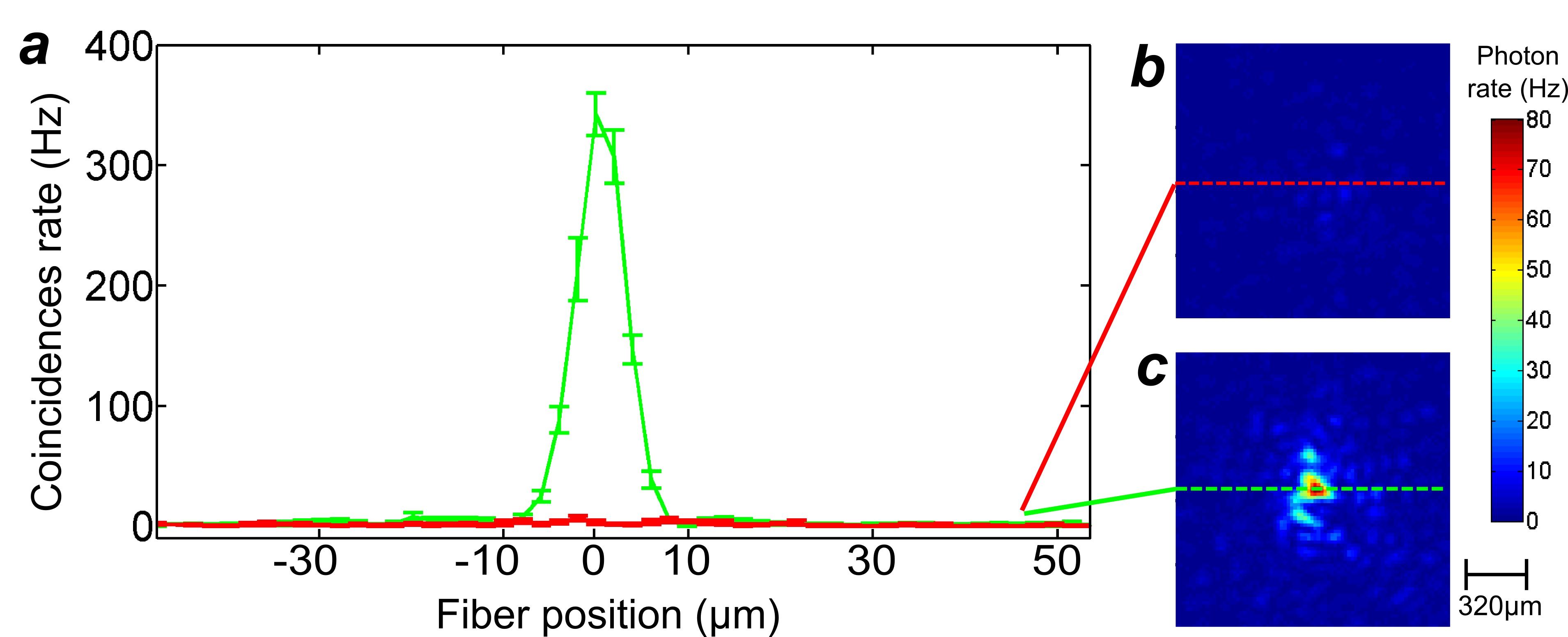}
\caption{\label{fig:focusing} Experimental results of single-photon focusing. (a) 1D scans of the output PMSMF realized with a random phase pattern (red curve) and the focusing phase pattern (green curve) displayed on the SLM. (b and c) Output plane images recorded with the EMCCD camera for the same random phase pattern (b) and the focusing pattern (c) displayed on the SLM. Red and green dashed lines represent the position of the scans realized with the output fiber. Contrary to the fiber output signal, measured in coincidences with the trigger photon of the non-classical source, intensity images recorded with the EMCCD camera are not triggered.}
\end{figure}

By displaying the corresponding phase pattern on the SLM, calculated using the SM, the non-classical light can be controlled and focused into any output modes of the medium. Results of single-photon focusing obtained are shown in Fig.~\ref{fig:focusing}. After positioning one output fiber in an arbitrary output position of the medium, two 1D fiber scans are performed. One scan is realized with a random phase pattern displayed on the SLM, and the other with the optimized phase pattern for focusing. The fiber output is measured in coincidences with the trigger photon of the non-classical source using a field-programmable gate array (FPGA) coincidence counter. The coincidence window time is set to $2.5$ ns. This method concentrates approximately $7 \%$ of the total output coincidence rate into the selected output PMSMFs. The total output coincidences rate is evaluated by multiplying the coincidence rate measured directly at the output of the pair-generation setup by the transmission coefficient of the wavefront shaping setup. Although focusing of a single-photon in a selected output fiber has been realized using an optimization technique~\cite{Huisman13}, here we demonstrate this without need for measurement feedback.

Due to space constraints, the two output PMSMFs could not be positioned next to each other in the same output plane. Consequently, each fiber is positioned in two conjugate output planes and a combination of an HWP ($\theta \simeq 45°$) and a PBS (Fig.~\ref{fig:setup}) is used to distribute equally the intensity between two arms. We denote by [A,B] the two output spatial modes of the medium locally injected in the two output PMSMFs. By focusing a single-photon into one of the two PMSMFs using the SM method, we generate single-photon states of the type $|01\rangle_{AB}$ or $|10\rangle_{AB}$. Because both fibers can be moved to any positions in the output plane, output modes [A,B] are arbitrary chosen  within the diffusive halo. 

\begin{figure}
\includegraphics[scale=0.9]{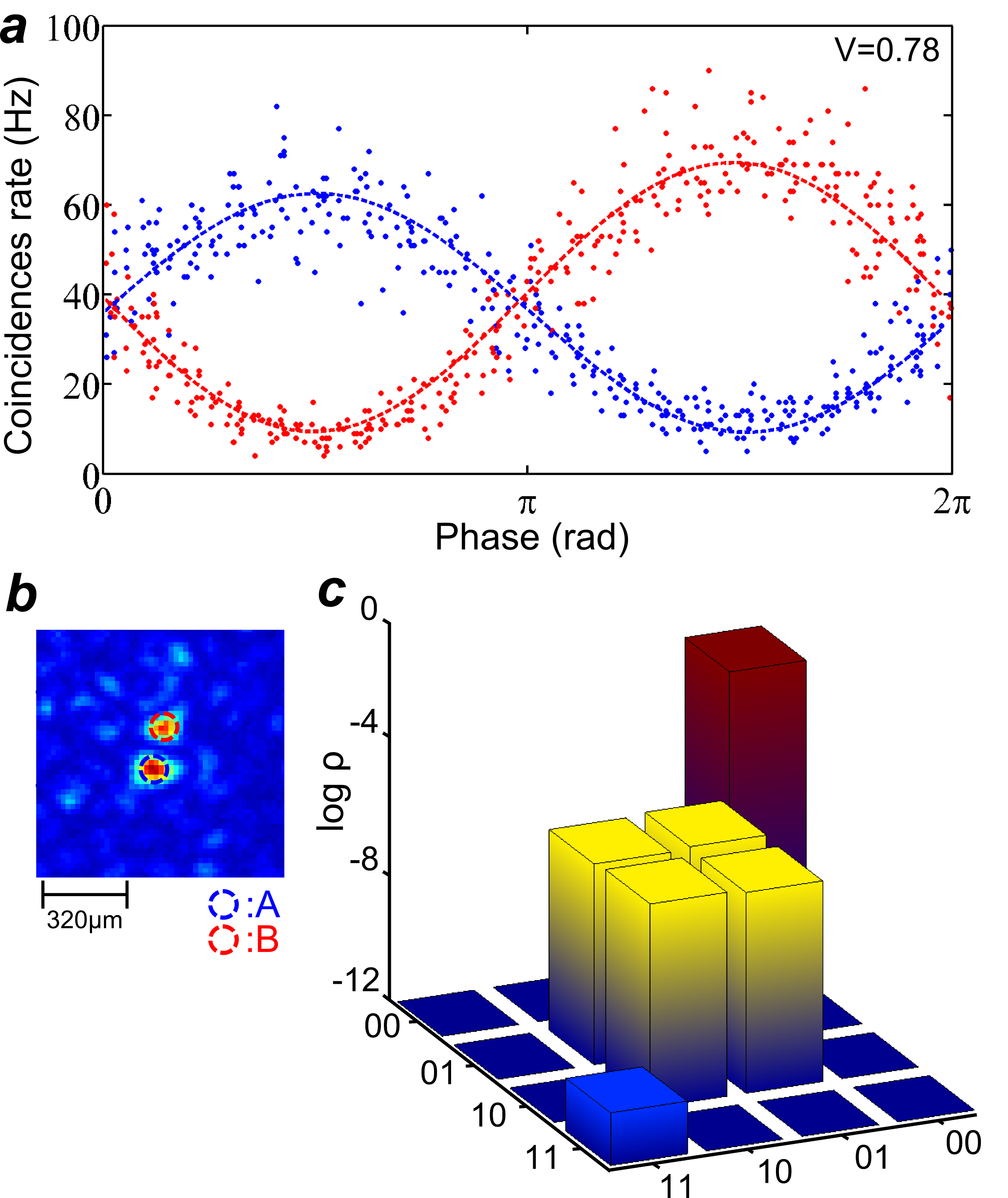}
\caption{\label{fig:densitymatrix} Experimental results of single-photon entangled state generation and characterisation. (a) Single-photon interference fringes observed by combining the two output PMSMFs on a fibered beam splitter (visibility $V=0.78 \pm 0.04$). The relative phase between the two output is scanned from $0$ to $2 \pi$ by displaying on the SLM a set of phase patterns calculated using the SM. Each phase pattern corresponds to an output field focused in both fibers simultaneously with a relative phase $\phi \in [0,2 \pi ]$. (b) Output plane image recorded with the EMCCD camera using the SLM phase pattern focusing in both output PMSMFs simultaneously. (c) Reduced density matrix of the generated state. Diagonal coefficients are evaluated measuring coincidences between the two outputs modes [A,B] and the trigger detector T. Off-diagonal coefficient is determined using the formula  $|d| \simeq V (P_{01}+P_{10})/2$. The density matrix elements are measured running a 3 hours acquisition with a coincidence window time of $2.5$ ns: $P_{00} = 1-8.4 \times 10^{-5}, P_{01} = (4.1 \pm 0.7) \times 10^{-5} ,P_{10} = (4.3 \pm 0.7) \times 10^{-5}, P_{11} = 1/(1.1 \times 10^{10}), |d| = (3.3 \pm 0.6) \times 10^{-5}$. From these data, a lower bound for the concurrence is calculated: $C = (4.6 \pm 1.2)\times 10^{-5}  > 0$. }
\end{figure}

In our experiment, we adopt the SM method to generate an entangled state of the form $\frac{\ket{01}_{AB}+e^{i \phi}\ket{10}_{AB}}{\sqrt{2}}$, by taking advantage of the fact that our device can address both PMSMFs simultaneously with a specific and controllable relative phase $\phi$ and amplitude. The positions of the two output modes [A,B] can be visualised in the image plane by the EMCCD camera (Fig.~\ref{fig:densitymatrix}b).

We now turn to investigate the coherence properties of the two modes; this is performed by investigating their reduced density matrix~\cite{Chou05,Lee11} 

\begin{equation}
\rho = \left( \begin{matrix}
P_{00} & 0 & 0 & 0 \\ 
0 & P_{01} & d & 0 \\
0 & d^* & P_{10} & 0 \\
0 & 0 & 0 &  P_{11} \\
\end{matrix} \right)
\end{equation}

where $P_{ij}$ represent the probability of finding $i$ photons in mode A, and $j$ in mode B. The $P_{ij}$ can be measured directly by photon counting on both modes. The coherence term is not directly accessible, but can be estimated by interfering the two modes.
We observe single-photon interference by combining the two output signals on a fiber beam splitter. To record interference fringes, represented on Fig.~\ref{fig:densitymatrix}a, a set of SLM phase patterns is first calculated using the SM. Each phase pattern corresponds to an output field focused into the two output PMSMFs simultaneously with equal amplitude and a relative phase  $\phi \in [0,2 \pi ]$. Then, we reconstruct the interference fringes by displaying successively on the SLM the different phase patterns with a relative phase varying from $0$ to $2 \pi$ in 21 steps. The interference visibility is evaluated to be $V=0.78 \pm 0.04$. This gives an estimation $|d|\simeq V(P_{10}+P_{01})/2 = (3.3 \pm 0.6) \times 10^{-5}$.

The level of entanglement in the state $\rho$ has a lower bound given by the concurrence \cite{Wootters98}. The concurrence, C, is a monotonic measure of bipartite entanglement that is zero for any separable state and positive for all entangled states. For this purpose, a reduced part of the density matrix $\rho$ is reconstructed in Fig.~\ref{fig:densitymatrix}c and the concurrence is calculated from the equation 

\begin{equation}
C = \max (2 |d| - 2 \sqrt{P_{00} P_{11}},0) 
\label{concurrence} 
\end{equation}

This value provides a strict lower bound to the entanglement. With no correction for detection efficiencies or propagation losses, and without substraction of any background, we find a lower bound for the concurrence: $C = (4.6 \pm 1.2) \times 10^{-5} >0$. 

To determine our confidence in concluding that the system is entangled, we estimate the Poissonian confidence level for positive concurrence. In our experiment, the number of triple coincidences $N_{ABT}$ measured during the density matrix reconstruction process has the most important impact on this confidence level. The density matrix represented in Fig.~\ref{fig:densitymatrix}c has been reconstructed during a 3-hour acquisition process where only one triple coincidence  was recorded, so that $N_{ABT}=1$. Given the measured trigger rate, the maximum value allowed in order that the concurrence remain positive is $N^0_{ABT}=11$ [35]. Therefore we have a confidence level of $99 \%$ that the measured concurrence is positive. This conclusively demonstrates a non-zero degree of entanglement in the generated single-photon state.

In conclusion, we have demonstrated that a multiple scattering system can be used to control non-classical states of light, in this case a single-photon Fock state. By characterizing of the scattering matrix of the medium and shaping the photonic wavefront we focus the single photon to an arbitrary output mode (speckle grain). By manipulating the single photon into two arbitrary orthogonal modes and performing tomography of the output quantum state, we unambiguously demonstrate the generation of a single-photon entangled state. This experiment shows that the control of the light in the scattering medium allows a coherent superposing of a non-classical state of light between different spatial modes in the scattering medium. 

This work which uses complete knowledge of the SM of a multiple scattering medium opens a number of potential routes to implement quantum information processing. While we have demonstrated control over two independent spatial modes,  the complexity of the scattering process could allow access to a significantly larger number of modes (easily on the order of $N = 10^4$), hence a much larger Hilbert space. The scattering process could therefore be seen as a complex mode converter, potentially allowing  the generation of  an N mode entangled state of the form $ \ket{W} = \ket{10..0} + e^{i \phi_2}  \ket{01..0} +...+e^{i \phi_N}  \ket{00..1}$, possibly  in a simpler and more scalable way than using interferometers \cite{Papp09} or photonic lattices \cite{Peruzzo10}. This experiment could also be extended towards control of broadband non-classical sources,  using spatio-temporal light shaping techniques \cite{Aulbach11,Katz11}. Even more intriguing there is the potential to control states with  several indistinguishable photons propagating through the medium, which would allow to observe and control multi-photon speckle~\cite{Spring13,Peeters10}.

The authors thank Lijian Zhang, Michael Sprague and Christoph S\"{o}ller for fruitful discussions and David Martina for technical support. This work was funded by the European Research Council (grant no. 278025) and partially supported by the UK EPSRC  (EP/H03031X/1 and EP/K034480/1).

\end{document}